# Ordered Growth of Topological Insulator $Bi_2Se_3$ Thin Films on Dielectric Amorphous $SiO_2$ by MBE†

Sahng-Kyoon Jerng,[a] Kisu Joo,[b] Youngwook Kim,[c] Sang-Moon Yoon,[d] Jae Hong Lee,[a] Miyoung Kim,[d] Jun Sung Kim,[c] Euijoon Yoon,[b] Seung-Hyun Chun,[a] and Yong Seung Kim[a,*]

[a] Department of Physics and Graphene Research Institute, Sejong University, Seoul 143-747, Korea

[b] Department of Nano Science and Technology, Graduate School of Convergence Science and Technology, Seoul National University, Suwon 443-270, Korea

[c] Department of Physics, Pohang University of Science and Technology, Pohang 790-784, Korea

[d] Department of Materials Science and Engineering, Seoul National University, Seoul 151-744, Korea

†Electronic supplementary information (ESI) is available: Influence of Se passivation, RHEED patterns during the growth and annealing procedures, RHEED pattern and XRD profile of epitaxially grown $Bi_2Se_3$ films on $Al_2O_3(0001)$ substrate, additional HRTEM images, AFM images, and Hall effect measurements.

* Corresponding Author: FAX: +82 2 3408 4316. E-mail address: yongskim77@gmail.com (Y. S. Kim)

**Abstract**

Topological insulators (TIs) are exotic materials which have topologically protected states on the surface due to the strong spin-orbit coupling. However, a lack of ordered growth of TI thin films on amorphous dielectrics and/or insulators presents a challenge for applications of TI-junctions. We report the growth of topological insulator $Bi_2Se_3$ thin films on amorphous $SiO_2$ by molecular beam epitaxy (MBE). To achieve the ordered growth of $Bi_2Se_3$ on amorphous surface, the formation of other phases at the interface is suppressed by Se passivation. Structural characterizations reveal that $Bi_2Se_3$ films are grown along the [001] direction with a good periodicity by van der Waals epitaxy mechanism. Weak anti-localization effect of $Bi_2Se_3$ films grown on amorphous $SiO_2$ shows modulated electrical property by the gating response. Our approach for ordered growth of $Bi_2Se_3$ on amorphous dielectric surface presents considerable advantages for TI-junctions with amorphous insulator or dielectric thin films.

## Introduction

Topological insulators (TIs) are recently emerging new class of materials which have topologically protected states on the surface.[1] Due to the strong spin-orbit coupling, the surface states exhibit distinct features such as Dirac linear energy dispersion inside the bulk gap, spin-polarization by spin-momentum locking nature, and weak anti-localization (WAL).[1-5] These physical phenomena have created immense interests in the past few years. Among chalcogenide compounds, $Bi_2Se_3$ has attracted a special attention as three-dimensional TIs because it has the largest bulk band gap of 0.3 eV and a well-defined single Dirac cone at the momentum zero point in k space.[2, 5] So far, mechanical exfoliations from single crystals[6, 7] have been widely used to obtain high quality $Bi_2Se_3$ on dielectric substrates, which show limited lateral sizes of only few micrometers with irregular shapes. However, for practical applications of TIs, the growth of large-scale TI thin films on dielectrics or insulators is a challenge to advance forward. For example, low-power spintronic applications could be facilitated by developing TI tunnel junctions with dielectric or ferromagnetic insulators.[8] Also, multichannel Dirac fermions could be utilized in spintronics by inserting normal insulators between multiple TI layers.[9, 10]

Since the widely used dielectrics or insulators (such as $SiO_2$, $HfO_2$, and $Al_2O_3$) for device applications are formed in amorphous states, the growth of $Bi_2Se_3$ thin films on amorphous surface is urgently needed for TI study, which can be appealing for both fundamental physics and device applications. Recently, vapor-solid synthesis method is utilized to grow few-layer nanoplates of $Bi_2Se_3$ on amorphous dielectric,[11] but the lateral sizes did not exceed several micrometers. On the other hand, molecular beam epitaxy (MBE) technique has an advantage to grow large-area $Bi_2Se_3$ thin films with controllability of thickness.[12,13] To date, however, only crystalline substrates (such as Si, GaAs, $Al_2O_3$, etc.) have been used to grow $Bi_2Se_3$ films on them.[12-24] In this paper, we report an approach to grow large-scale $Bi_2Se_3$ thin films on amorphous $SiO_2$ by MBE. As a substrate, we chose 300 nm thick amorphous $SiO_2$ on Si(100), which had been widely used for bottom-gating experiments of $Bi_2Se_3$ flakes.[6, 7, 11] At the initial stage of epitaxial growth, the formation of crystalline $Bi_2Se_3$ was hindered by the dangling bonds of surface defects, which was observed by *in situ* reflection high-energy electron diffraction (RHEED). However, by applying Se passivation to the surface, we were able to grow c-axis oriented $Bi_2Se_3$ films on amorphous $SiO_2$ with the aid of strain relaxation by van der Waals epitaxy mechanism. Structural characterizations by high-resolution transmission electron microscopy (HRTEM) show a periodicity of $Bi_2Se_3$ film along the [001] direction. Even though $Bi_2Se_3$ films were grown on amorphous surface, electrical transport measurements show the typically observed WAL effects in TI films and can be further modulated via bottom-gating. This growth compatibility of $Bi_2Se_3$ and amorphous dielectrics will stimulate the fabrication of TI hetero-junctions for practical applications.

## Experimental section

### Growth

$Bi_2Se_3$ thin films were grown by a home-made MBE system which was equipped with Knudsen cells and high purity bismuth (99.999+%) and selenium (99.999%) sources. The working pressure was less than $2.0 \times 10^{-9}$ Torr with liquid nitrogen flowing in the shroud. $SiO_2$ substrates, cleaned by acetone and isopropyl alcohol prior to $Bi_2Se_3$ growth, were heated up to 120 °C for 30 minutes by a tungsten filament behind the sample manipulator in ultrahigh vacuum chamber to remove residual contaminations. The passivation process of surface dangling bonds was conducted by exposing the $SiO_2$ substrate to Se flux at 120 °C (this temperature is low enough to prevent the formation of second phases such as $SiSe_2$ and the Se layer could be self-limited without forming thick Se layer at this temperature).[14] The substrate temperature was increased to 250 °C for the growth of $Bi_2Se_3$ thin films. A quartz crystal microbalance thickness monitor (Inficon SQM160) was used to calibrate the ratio of Bi and Se fluxes. The deposition rate of $Bi_2Se_3$ films was ~0.67 Å/min in this study. The ratio of Bi and Se flux was kept at ~1:15 to minimize Se vacancies.[12, 21, 25] The formation of crystalline $Bi_2Se_3$ on the substrate is monitored by *in situ* RHEED pattern. For electrical transport measurements, samples were capped by Se layers on top after $Bi_2Se_3$ growth in order to minimize the contamination and/or doping effect from air exposure.[26]

**Characterization**

Raman spectroscopy measurement was carried out using a micro-Raman spectroscopy system (Renishaw inVia) with a 514.5 nm laser excitation source. The incident laser had a power of 3.75 μW and was focused through an 100x objective. The spectra were calibrated by the Si peak at 520 $cm^{-1}$. X-ray diffraction (XRD) patterns were obtained using a X-ray diffractometer (Bruker D8 Advance) with Cu Kα radiation of wavelength 1.5406 Å. Interface studies were carried out using a HRTEM (Tecnai F20) operated at 200 kV. TEM samples were prepared by standard mechanical polishing. Surface morphology of the samples was evaluated by a commercial atomic force microscopy (AFM, PSIA XEI100). Low-temperature transport properties were measured by using the standard six-probe method in physical property measurement system (PPMS, Quantum Design).

**Results and discussion**

The sequence of the layered $Bi_2Se_3$ is Se-Bi-Se-Bi-Se which called a quintuple layer (QL, corresponding to ~0.955 nm) as shown in Fig. 1a. The surface is terminated by Se atoms without dangling bonds, resulting in only weak van der Waals (vdW) forces between each QL. Due to a weak interaction of overlayer with the substrate, $Bi_2Se_3$ can be grown epitaxially even on highly lattice mismatched surface, which is called van der Waals epitaxy (vdWE).[27, 28] We tried to grow crystalline $Bi_2Se_3$ thin films on amorphous $SiO_2$ via vdWE, and found that the crystallinity of $Bi_2Se_3$ thin film could be further enhanced by Se passivation of surface dangling bonds on $SiO_2$ substrate. As shown in Fig. 1b, the RHEED pattern of $SiO_2$ is a haze due to the amorphous surface. When the $Bi_2Se_3$ film is grown on unpassivated $SiO_2$, the RHEED pattern changes to the spotty patterns (marked by white arrows in Fig. 1c) which indicate that the growth of $Bi_2Se_3$ is encumbered by a formation of other phases (see also the ESI). On the other hand, a streaky RHEED pattern of crystalline $Bi_2Se_3$ is observed when the growth proceeds on Se passivated $SiO_2$ surface (Fig. 1d), indicating enhanced crystallinity of $Bi_2Se_3$. Once the surface dangling bonds are saturated, chemical bond formation between the substrate and TI components can be prevented.[14] Additionally, enlarged grain sizes of $Bi_2Se_3$ at the initial stage are observed on Se passivated $SiO_2$ (see AFM images in the ESI). Fig. 1e shows a photograph of $Bi_2Se_3$ film grown on 1.5 x 1.5 $cm^2$ substrate (edges of substrate were masked by a Mo holder).

To investigate the film crystallinity, as-grown 20 nm thick $Bi_2Se_3$ films on amorphous $SiO_2$ were identified by XRD. Fig. 2a shows that the XRD profile contains only the (003) family of diffraction peaks in the $\theta$-$2\theta$ scan, whose positions agree well with those of c-axis oriented epitaxial $Bi_2Se_3$ films (see the ESI) and bulk values.[12] Fig. 2b shows the vibrational modes and the Raman spectra in the range of 100—200 $cm^{-1}$. Within these frequencies, two characteristic peaks were observed at ~131 $cm^{-1}$ and ~174 $cm^{-1}$, which correspond to an in-plane mode ($E_g^2$) and an out-of-plane mode ($A_{1g}^2$) of rhombohedral $Bi_2Se_3$ lattice vibrations, respectively.[16, 29-31]

For detailed structural information, the cross-sectional HRTEM was used to examine the microstructure of $Bi_2Se_3$ film grown on $SiO_2$ substrate. Fig. 3a and 3b show the HRTEM images of periodic atomic structure of $Bi_2Se_3$ on amorphous $SiO_2$ substrate. As marked by orange colored arrows in Fig. 3a, some structural faults are observed in $Bi_2Se_3$ film, which are similarly observed in a recent study of MBE grown $Bi_2Se_3$ films on highly mismatched crystalline substrates.[18] This result can be attributed to the slightly wavy surface of substrate and the surface roughness of substrate. Note that the difference of surface potentials, correlated with roughness such as point defects, affects the initial growth of $Bi_2Se_3$ films on amorphous substrate. As shown in Fig. 3b, initial 1—2 QLs (orange colored arrow) are not exactly parallel to the surface near the small puddle of $SiO_2$ surface. However, after the puddle is filled with initial few layers, the strain is released along the growth direction due to the vdWE mechanism. The HRTEM image of high magnification (Fig. 3b) clearly reveals the crystalline structure of $Bi_2Se_3$ along the [001] direction with the distance of 0.965 nm between QLs (white dot lines), which is consistent with the cross-sectional height (~0.96 nm) of triangular feature on the top surface (see AFM images in the ESI). These results show that our approach achieved the ordered growth of $Bi_2Se_3$ thin films on amorphous surface. Furthermore, we infer that other layered TI thin films of chalcogenide compounds can be grown on amorphous surface by using proper surface passivation technique such as *in situ* plasma treatment or exposure of chalcogen materials to saturate the dangling bonds.

Even though $Bi_2Se_3$ film was grown on amorphous surface, electrical transport measurements show typically observed WAL effects in TI films and can be further modulated via bottom-gating. As shown in the inset of Fig. 4a, the normalized magnetoresistance of 20 nm thick $Bi_2Se_3$ film shows a cusp at B = 0 T, which is a signature of WAL effect originated from quickly suppressed destructive quantum interferences between time-reversed paths in a perpendicular magnetic field.[32] This WAL behavior reflects both the Dirac nature of the surface states and the strong spin-orbit interaction in the bulk of TIs.[33] The exact behavior of magnetoconductance ΔG(B) is described by the Hikami-Larkin-Nagaoka formula:[34] $\Delta G(B) = \alpha(e^2/\pi h)[ln(B_\varphi/B)-\Psi(1/2+B_\varphi/B)]$, where ΔG(B) is the low field magneto-conductance (ΔG(B) =

G(B)-G(0)), $B_\varphi$ is the dephasing magnetic field, Ψ is the digamma function, and α is a coefficient predicted to be 1/2 for each surface channel. Fig. 4a presents the magnetoconductance at various gate voltages ($V_g$); the gate contact was simply made to the bottom of p-type Si substrate using amorphous $SiO_2$ as dielectrics. The obtained value of α is close to 1/2 (~0.43 at $V_g$ = –50 V), shown in upper panel of Fig. 4b, implying that the top and bottom surfaces are coupled through the bulk to form a single effective channel for the phase-coherent transport.[10, 11, 35] Note that the coefficient α decreases from 0.43 to 0.39 (upper panel of Fig. 4b) while the carrier concentration increases from 3.4 to 4.8 × $10^{13}$ $cm^{-2}$ (lower panel of Fig. 4b, extracted from the Hall measurements, see also the ESI) as the $V_g$ increases from –50 to 50 V. This is probably due to the fact that the bulk carriers contribute more to the overall conductivity of the system with the increasing gate voltage.[35] Therefore, further improvements should be possible by reducing the bulk carrier concentration through Ca doping or gating with a high-k dielectric layer.[6, 20, 33] The existence of the WAL effect with the gating response bears a considerable potential in the application of our approach for TI-junctions of $Bi_2Se_3$ films and dielectrics or insulators which are formed in amorphous states.

## Conclusions

In conclusion, we demonstrate the ordered growth of $Bi_2Se_3$ thin films on the dielectric amorphous $SiO_2$ surface by vdWE. Passivation of the dangling bonds was accomplished by low-temperature Se exposure. Structural characterizations show that c-axis oriented $Bi_2Se_3$ films were grown on $SiO_2$ surface via vdW coupling. Low-temperature transport measurements show the WAL effect and bottom-gating responses, which suggest that $Bi_2Se_3$ films directly grown on amorphous $SiO_2$ exhibit the surface state properties of TI materials. Our results will stimulate the application of TI materials in combination with amorphous dielectrics.

**Acknowledgements**

This research was supported by the Priority Research Centers Program (2012-0005859), the Basic Science Research Program (2012-0007298, 2012-040278, 2012-013838), the Center for Topological Matter in POSTECH (2012-0009194), and the Nanomaterial Technology Development Program (2012M3A7B4049888) through the National Research Foundation of Korea (NRF) funded by the Ministry of Education, Science and Technology (MEST).

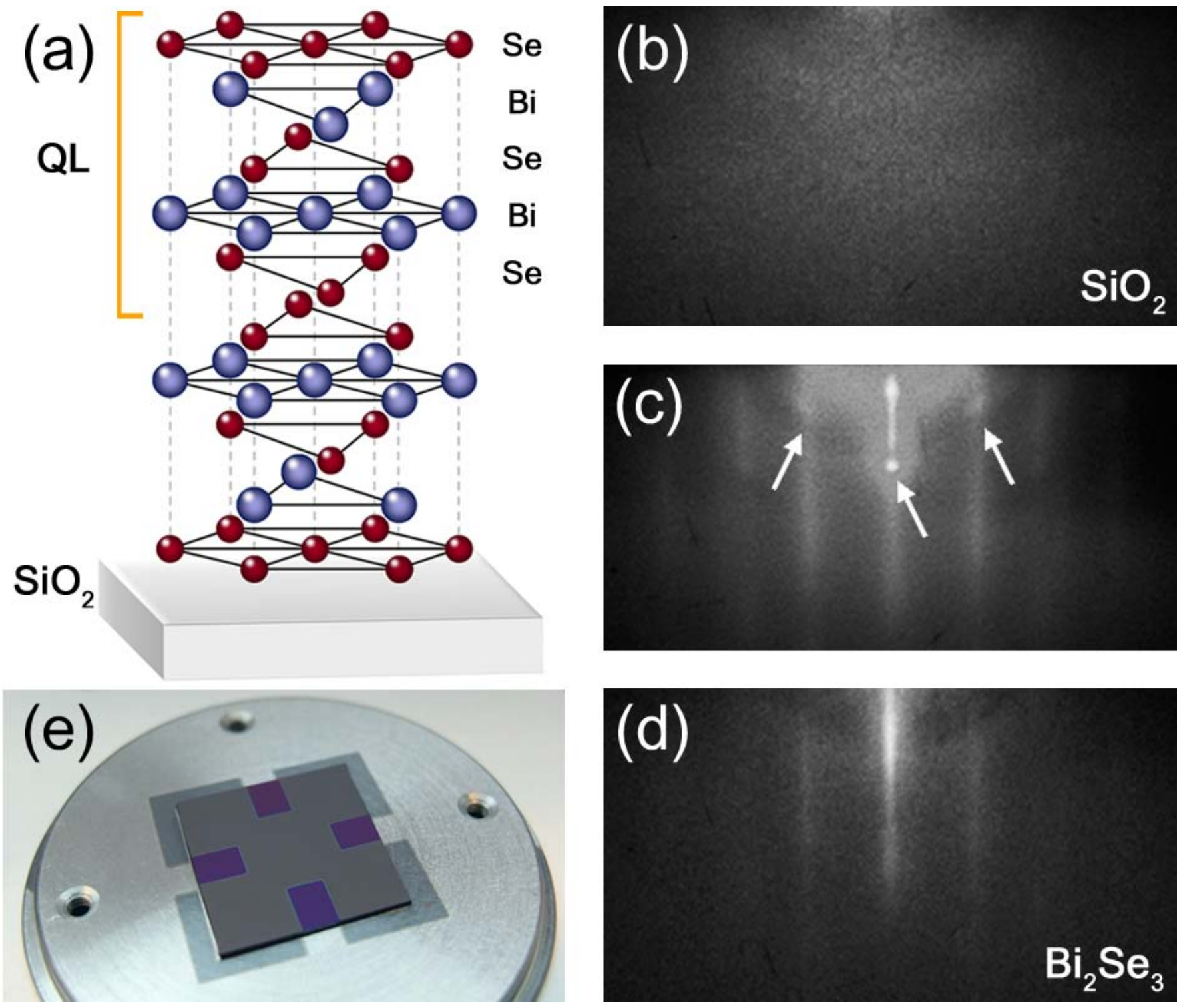

**Fig. 1** (a) Schematic illustration of $Bi_2Se_3$ structure on $SiO_2$ substrate. RHEED patterns of (b) amorphous $SiO_2$ substrate, (c) other phase of $Bi_2Se_3$ grown on unpassivated $SiO_2$, and (d) crystalline $Bi_2Se_3$ grown on Se passivated $SiO_2$. (e) Photograph of $Bi_2Se_3/SiO_2$ specimen (1.5 cm x 1.5 cm).

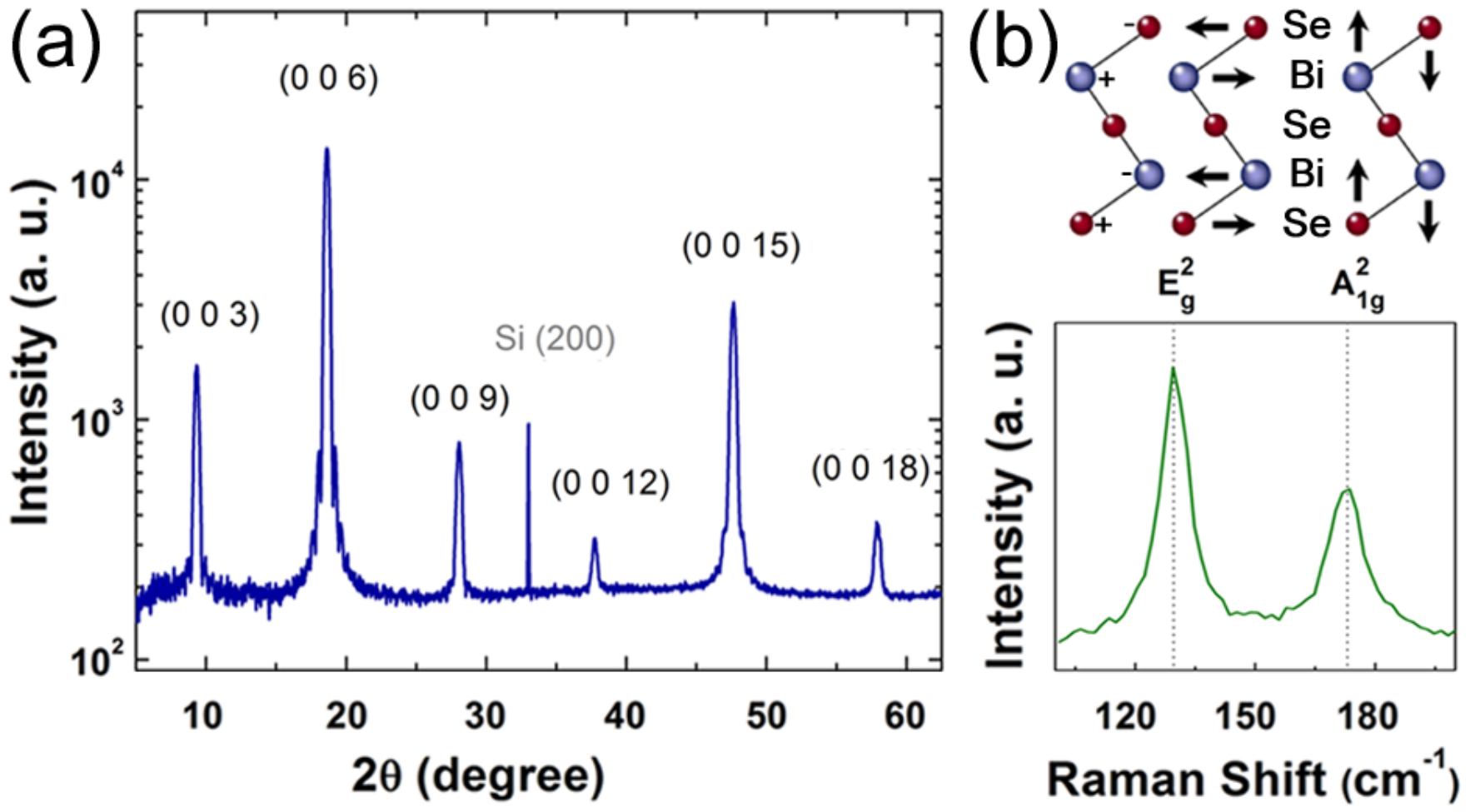

**Fig. 2** The structural characterizations of 20 nm thick $Bi_2Se_3$ thin films. (a) XRD pattern in the $\theta$-$2\theta$ scan. (b) Schematic of vibrational modes and corresponding Raman spectra.

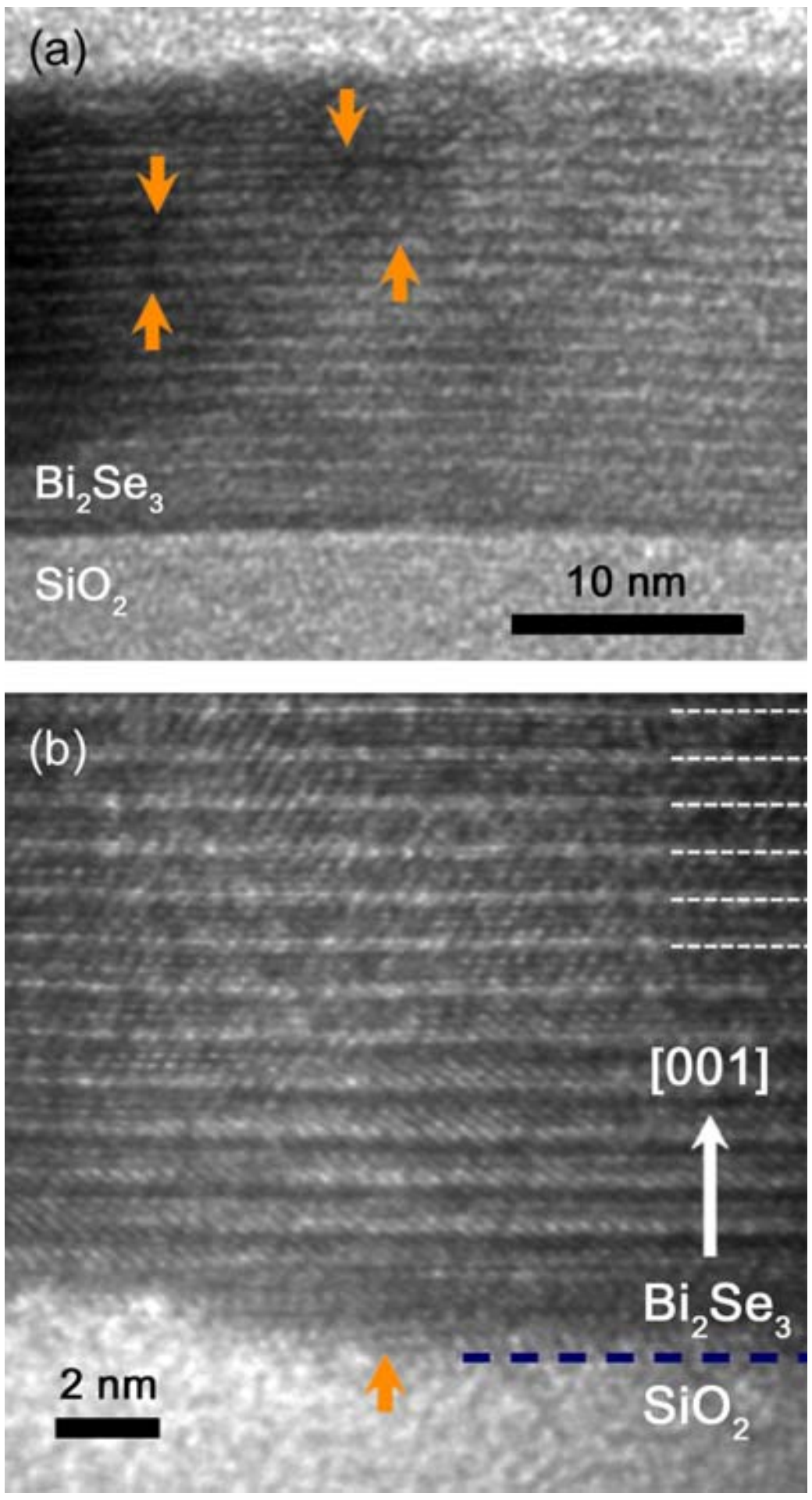

**Fig. 3** Cross-sectional HRTEM images at (a) low magnification and (b) high magnification. Some structural faults are observed within $Bi_2Se_3$ and at the interface with $SiO_2$ (pointed by the orange colored arrows). The dotted lines indicate QLs.

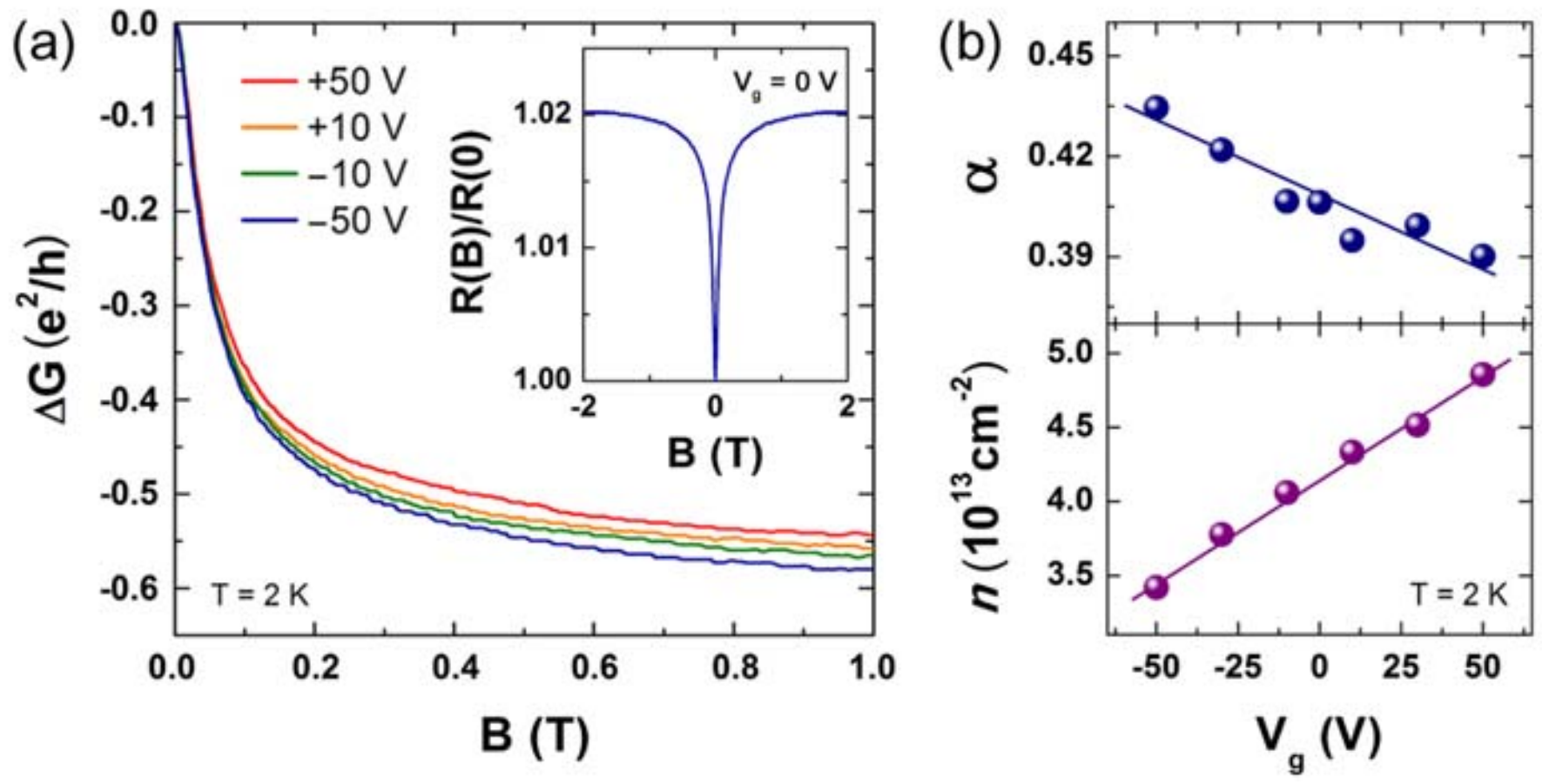

**Fig. 4** Electrical transport measurements of a 20 nm thick $Bi_2Se_3$ film at 2 K. (a) Gate voltage dependence of the WAL effect and (inset) normalized magnetoresistance at zero gate voltage. (b) The fitting coefficient α from Hikami-Larkin-Nagaoka theory (upper panel) and gate voltage dependence of the areal carrier concentration (lower panel).

**TOC**

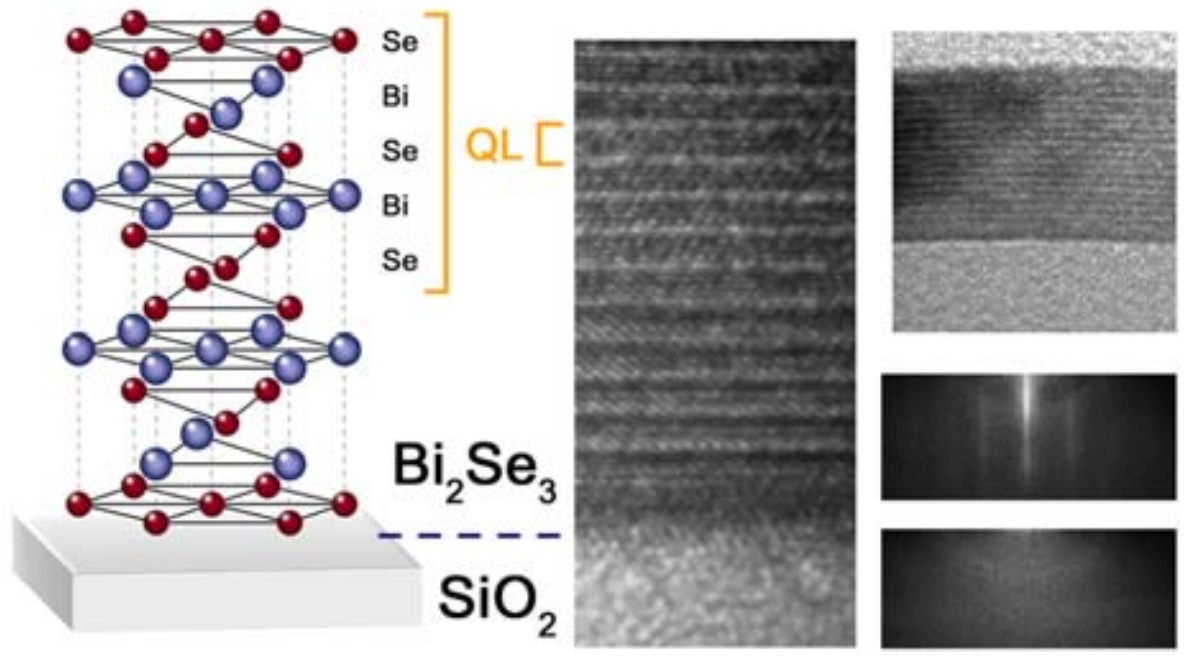

Topological insulator $Bi_2Se_3$ thin films are grown directly on oxidized amorphous silicon ($SiO_2$) substrate by molecular beam epitaxy with van der Waals epitaxy method.

# Supplementary Information

## Ordered Growth of Topological Insulator $Bi_2Se_3$ Thin Films on Dielectric amorphous $SiO_2$ by MBE

*Sahng-Kyoon Jerng, Yong Seung Kim, Kisu Joo, Youngwook Kim, Sang-Moon Yoon, Jae Hong Lee, Miyoung Kim, Jun Sung Kim, Euijoon Yoon, and Seung-Hyun Chun*

### 1. The influence of Se passivation

Fig. S1a and b show AFM images of $Bi_2Se_3$ films grown on unpassivated $SiO_2$ surface and Se passivatied $SiO_2$ surface, respectively. Surface morphology is obtained by commercial AFM (NanoFocus Inc). Both of samples were not annealed after growth of 2 hours. The morphology of as-grown films without passivation is exhibited small grains (Fig. S1a) and RHEED pattern was indicated a formation of other phase as shown in inset of Fig. S1a (marked by white arrow). On the other hand, as-grown films on Se passivated $SiO_2$ presents forming of large grains (Fig. S1b) and RHEED pattern of well-crystallized $Bi_2Se_3$ (the inset of Fig. S1b). The surface roughness is 0.54 nm and 0.66 nm for $Bi_2Se_3$ films grown on the unpassivated surface and the passivated surface, respectively.

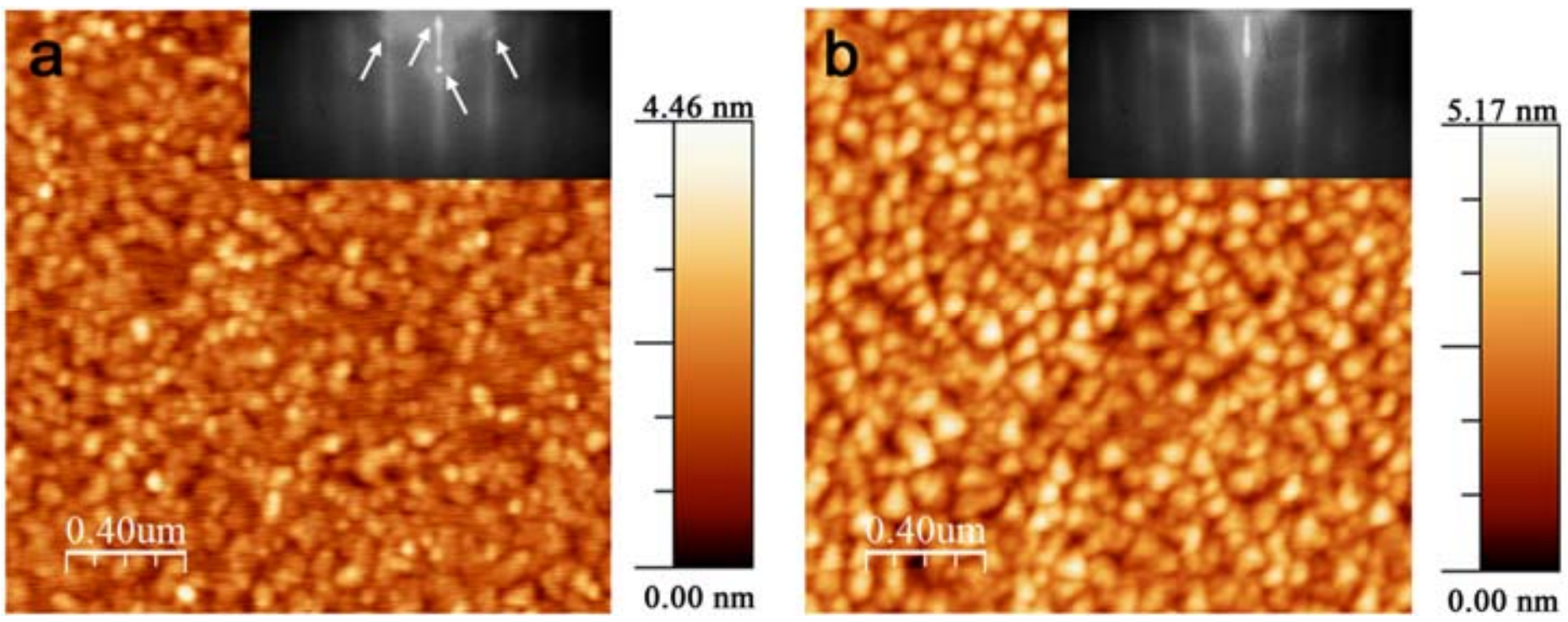

**Fig. S1** AFM images of $Bi_2Se_3$ films for (a) grown on unpassivated $SiO_2$ surface and (b) Se passivated $SiO_2$ surface. The insets are RHEED patterns recorded during the growth.

**2. RHEED patterns for $Bi_2Se_3$ growth procedures**

Prior to the growth, samples were heated at 120 °C to remove residual contamination. Fig. S2a shows RHEED pattern of SiO2 substrate at 120 °C, which is typical RHEED pattern for amorphous substrate. After 30 minutes, Se was deposited on $SiO_2$ surface. RHEED pattern was not changed from that of amorphous (Fig. S2b). Then, temperature increased to 250 °C for growth of $Bi_2Se_3$ films. During the growth, RHEED pattern is developed to streaky diffraction pattern as shown in Fig. S2c and S2d. After the growth, samples were annealed at 450 °C (Fig. S2e). RHEED patterns are not shown the formation of other phase in the annealing to cool down (Fig. S2f).

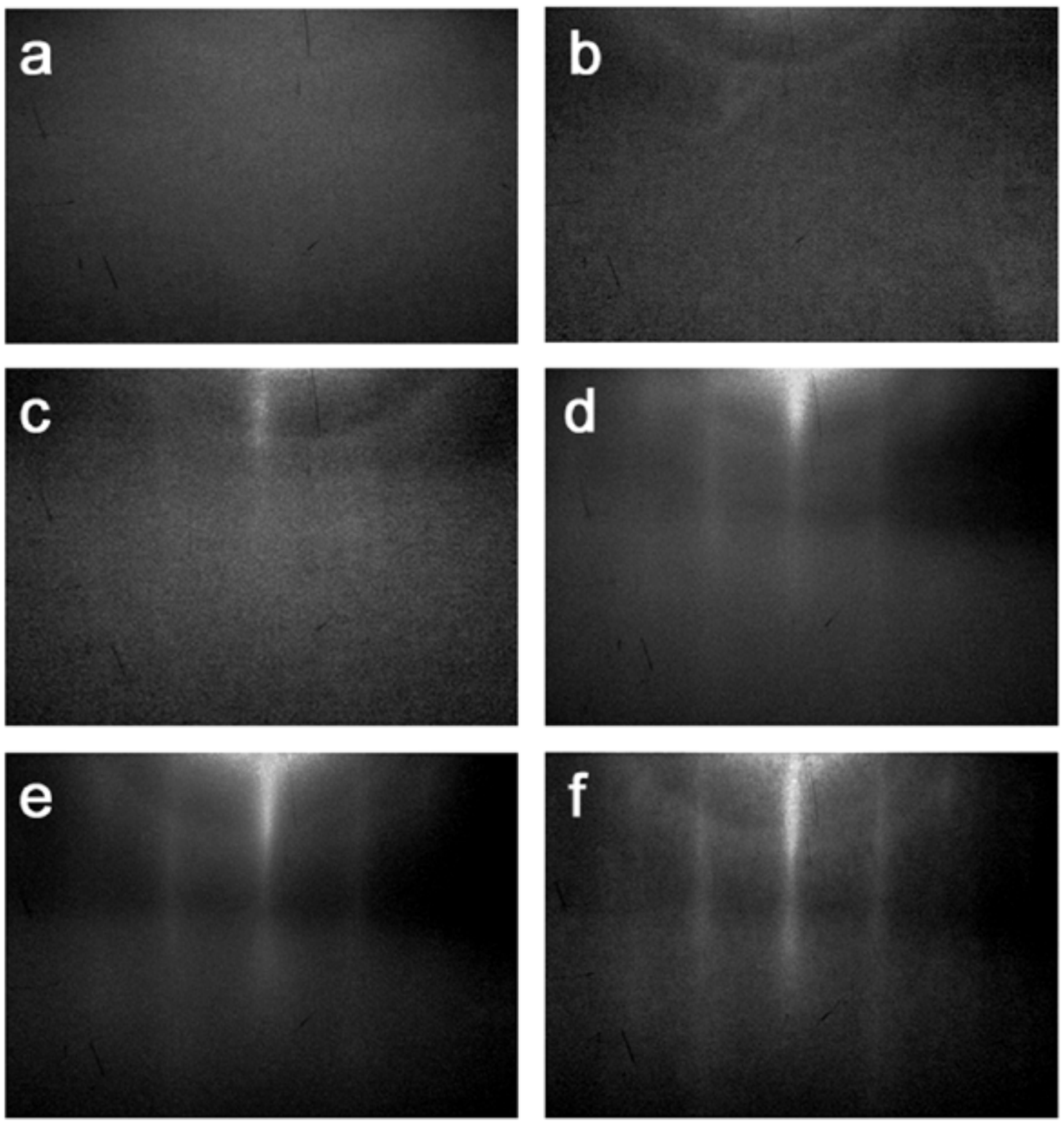

**Fig. S2** RHEED patterns for $Bi_2Se_3$ growth procedures: (a) $SiO_2$ substrate at 120 °C, (b) Se exposure for passivation, (c)-(d) $Bi_2Se_3$ growth (RHEED recorded at 30 min and 90 min, respectively), (e) annealing at 450 °C for 30 min, and (f) cool down. The large dark features on the screen are shadows of the substrate holders.

## 3. RHEED pattern and XRD profile of epitaxially grown $Bi_2Se_3$ films on $Al_2O_3$(0001) substrate

Fig. S3 provides RHEED patterns and XRD profile of MBE-grown epitaxial $Bi_2Se_3$ films on $Al_2O_3$(0001), known as sapphire, substrate. As shown in Figure S3a, RHEED pattern indicates crystalline sapphire substrate. During the growth of $Bi_2Se_3$ at 250 °C, RHEED pattern changes to the crystalline $Bi_2Se_3$. XRD profile, obtained by Shimadzu XRD-6100, reveals the (003) family of diffraction peak of c-axis oriented $Bi_2Se_3$ films as shown in Fig. S3c.

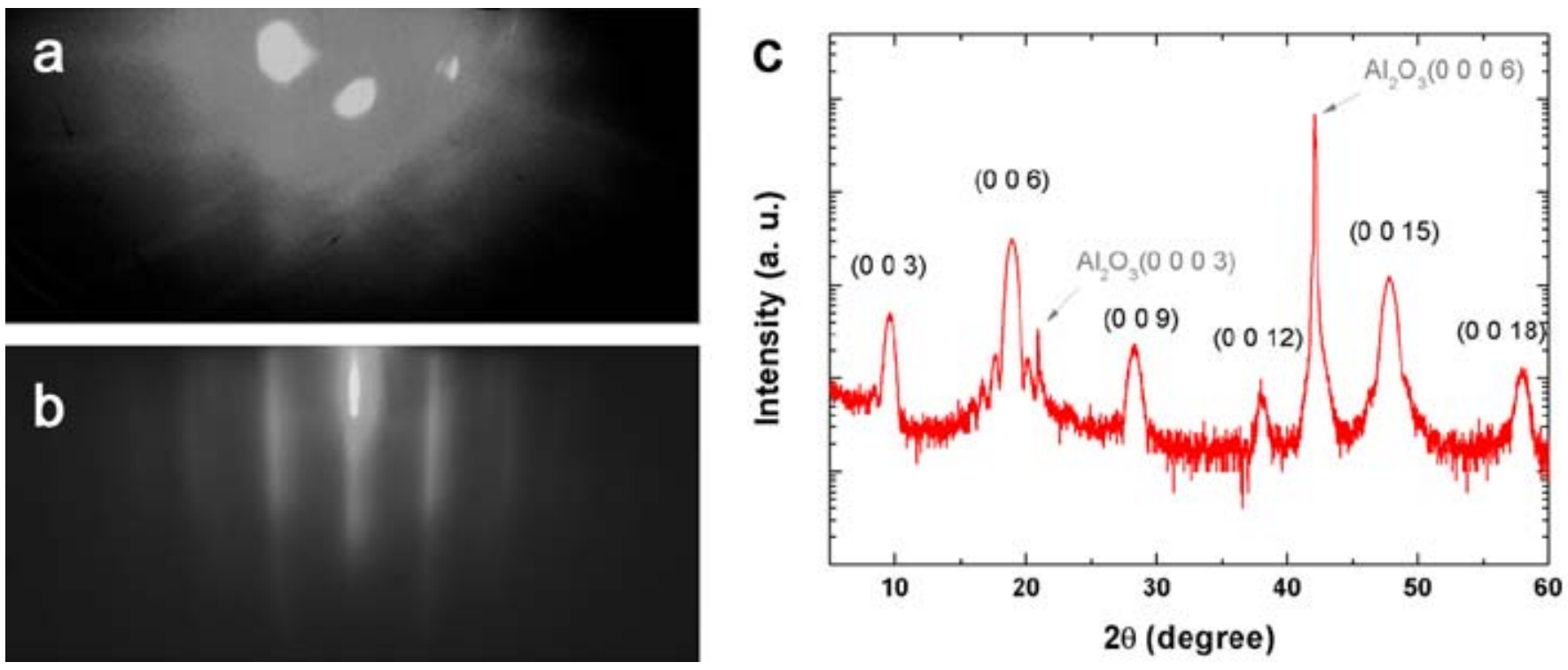

**Fig. S3** RHEED pattern of (a) crystalline sapphire substrate and (b) epitaxially grown $Bi_2Se_3$ films. (c) XRD profile of MBE-grown $Bi_2Se_3$ films on sapphire substrate.

## 4. Additional HRTEM images of interface between $Bi_2Se_3$ films and flat $SiO_2$

Fig. S4 shows the HRTEM images for different $Bi_2Se_3$ thin film samples grown on $SiO_2$. Additional HRTEM measurement was carried out using JEM-2100F instrument operated at 200 kV. Well-stacked $Bi_2Se_3$ quintuple layers are clearly observed.

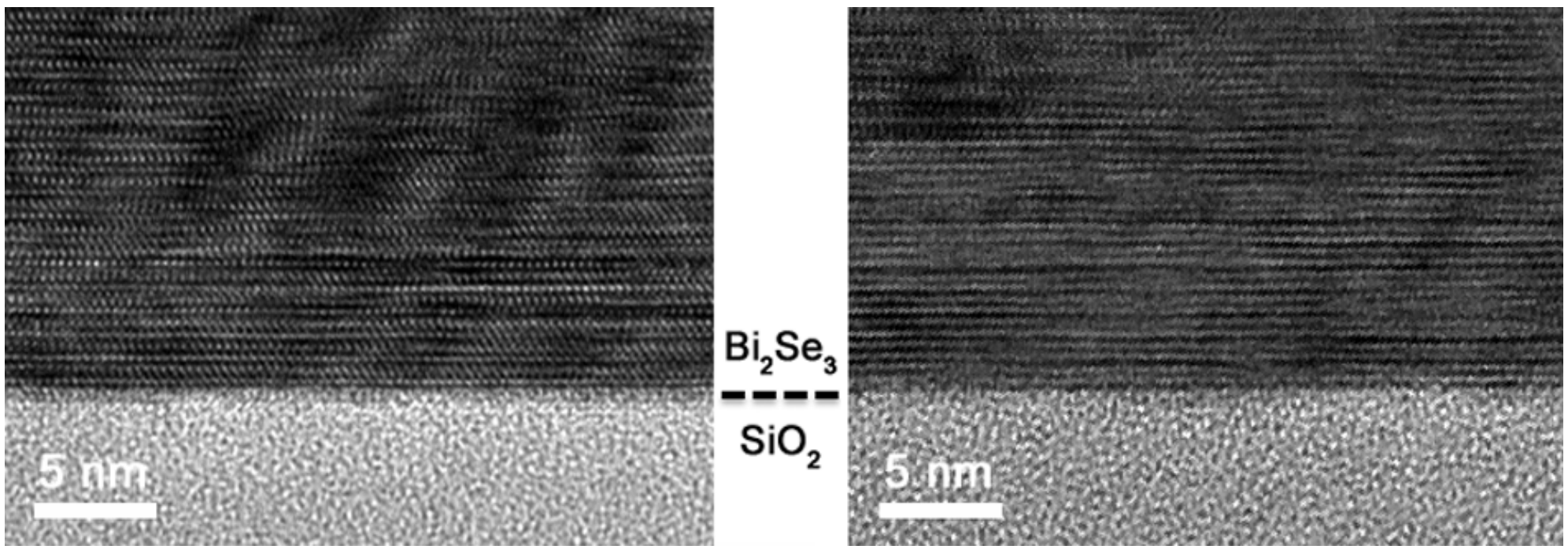

**Fig. S4** HRTEM images for $Bi_2Se_3$ thin films grown on flat $SiO_2$ surface area.

### 5. Surface morphology of $Bi_2Se_3$ films grown on $SiO_2$

Fig. S5 shows the AFM image of a 20 nm thick $Bi_2Se_3$ thin film grown on $SiO_2$. Terraces and/or steps (with triangular and hexagonal shapes) represent the 3-fold symmetry of $Bi_2Se_3$. The cross-sectional height of triangular feature is ~0.96 nm. The disordered feature of triangular $Bi_2Se_3$ terraces along the in-plane directions can be attributed to the 300 nm thick amorphous $SiO_2$ surface which does not provide a lattice constraint for atom stacking during the deposition.

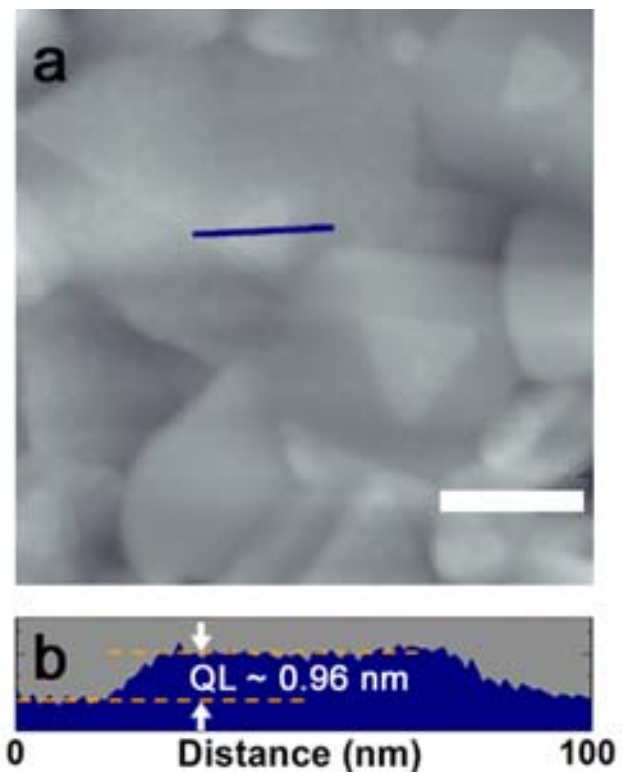

**Fig. S5** (a) AFM image of a 20 nm thick film. Scale bar = 100 nm. (b) Cross-sectional profile along the line in (a) showing the QL of $Bi_2Se_3$.

### 6. Hall measurement for 20 nm thick films

Fig. S6 shows Hall resistance of 20 nm thick films in the perpendicular magnetic field. The standard six-probe method is used for Hall measurement (I = 100 μA, T = 2 K). Sample is indicated as n-type semiconductor with a carrier concentration of 4.2 x $10^{13}$ $cm^{-2}$ at $V_g$ = 0 V. The carrier density decreased down to 3.4 x $10^{13}$ $cm^{-2}$ at $V_g$ = -50 V.

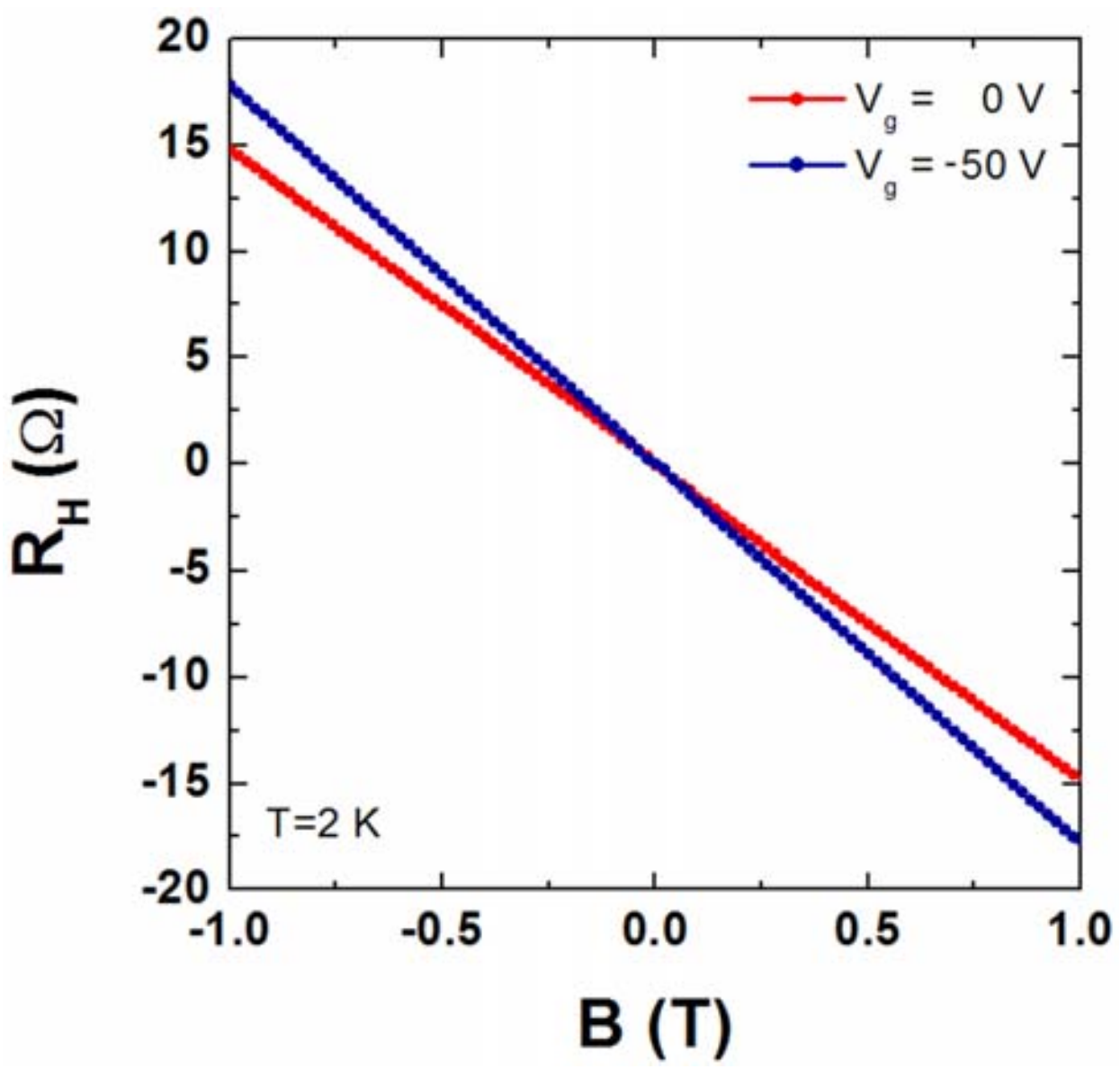

**Fig. S6** Hall resistance of 20 nm thick films for applying gate voltage in the magnetic field of ±1 T.